\documentclass[prl,twocolumn]{revtex4}
\usepackage{amssymb}

\input{tcilatex}

\begin{document}

\title{Far-Field Electron Spectroscopy of Nanoparticles }

\begin{abstract}
A new excitation mechanism of nanoparticles by relativistic, highly focused
electron beams, is predicted under nanoparticle recoil. The corresponding
electron energy loss spectra, calculated for metallic (silver and gold) and
insulating (SiO$_{2}$) nanoplatelets, reveal dramatic enhancements of
radiative electromagnetic modes within the light cone, allowed by the
breakdown of momentum conservation in the inelastic scattering event. These
modes can be accessed with e-beams in the vacuum far-field zone, offering
interesting capabilities for signal transmission across nano to meso length
scales.

PACS number(s): 79.20.Uv, 78.67.Bf, 73.20.Mf, 41.60.-m
\end{abstract}

\author{M. A. Itskovsky$^{1}$, H. Cohen$^{2}$ and T. Maniv$^{1}$}
\date{\today}
\affiliation{$^{1}$Schulich Faculty of Chemistry, Technion-IIT, 32000 Haifa, ISRAEL\\
$^{2}$Weizmann Institute of Science, Chemical Research Support, Rehovot
76100, ISRAEL }
\maketitle

A powerful approach to the study of nanoparticles is provided by very fast
(relativistic) electron beams (e-beams), with typical lateral resolution on
an atomic scale, available in scanning transmission electron microscopes
(STEM)\cite{Batson83,Howie85,Muller93,Cohen98,Lembrikov03,Cohen03}. As
discussed previously \cite{Batson83,Howie85,Cohen98}, when the e-beam is
restricted to the vacuum near a selected nanoparticle, its electromagnetic
(EM) interaction with surface plasmons or surface plasmon-polaritons (SPPs) 
\cite{Raether80} is reminiscent of the near-field interaction of
subwavelength optical probes. Several works have recently studied
realizations of Cherenkov radiation excitation within various dielectric
media by e-beams moving in near-field vacuum zones \cite%
{Zabala03,deAbajo98-99,Ochiai04,deAbajo03}.

In the present paper we show that by considering recoil effects of the
nanoparticle during the scattering event one introduces far-field coupling
between the electron and the nanoparticle, which dramatically enhances the
radiative channels in the loss spectrum. To illustrate our main points we
consider here a simple model where the e-beam is propagated in the vacuum
along a wide face of a rectangular nanoplatelet (oriented, e.g., in the $x-y$
plane), and a surface or guided wave induced by the electron is propagated
with a wave number $k_{x}$ along the beam axis. The spatially sensitive
nature of the corresponding electron energy loss process arises from the
exponential dependence, $e^{-2K^{\star }b}$, of the EM interaction between
the e-beam and the platelet on the impact parameter $b$. The extinction
coefficient, $K^{\star }=\sqrt{k^{2}-\left( \omega /c\right) ^{2}}$, with $%
k^{2}=k_{x}^{2}+k_{y}^{2}$ , determines the tail of the evanescent field in
the vacuum for values of $k$\ outside the light-cone, i.e. for $k>\omega /c$%
. Inside the light-cone, i.e. for $k<\omega /c$ , $K^{\star }$ is purely
imaginary and the corresponding interaction becomes spatially oscillating,
allowing the electron to exchange photons with the particle far away into
the vacuum. This striking possibility has been overlooked in the recent
literature of STEM-electron energy loss spectroscopy (EELS), since the
excitation by an electron moving in the vacuum with a classical velocity v,
has been restricted to a constant longitudinal wavenumber $k_{x}=\omega
/v>\omega /c$, implying EM coupling to the nanoparticle which is restricted
to the evanescent tail near the surface.

Following Ref.\cite{Cohen98}, the focused e-beam is described here as a
one-dimensional wave, propagating along the $x$-axis, while in the
transverse ($y-z$) directions it is described by a wave-packet localized
within a smoothly converging cross section along the beam axis, whose shape
is assumed to be squared for the sake of simplicity. The nanoplatelet
half-sides $a^{\star },b^{\star },c^{\star }$ along the $x,y,z$ axes
respectively, are assumed to satisfy $b^{\ast }\gg a^{\star }\gg c^{\star }$
(see below for more details). Within the framework of this
quantum-mechanical approach we use an effective beam-nanoparticle
interaction Hamiltonian based on the retarded four-vector image potential of
the e-beam and Born-Oppenheimer-like separation of coordinates (i.e. the
`slow' transverse coordinates from the `fast' longitudinal one), justified
by the small e-beam converging angle. The corresponding rate of change of
inelastic scattering probability of the incident e-beam at an impact
parameter $b$ and for energy loss $\Delta E=\hbar \omega $, calculated to
first order in the effective interaction Hamiltonian, $H_{EM}^{p_{x}}\approx
-e\Phi -e\frac{p_{x}}{mc}A_{x}$, can be written in the form:

\begin{eqnarray}
R &\propto &\sum_{\overrightarrow{q}_{tr}^{i},\overrightarrow{q}%
_{tr}^{f}}e^{-\beta \frac{\left( \hbar q_{tr}^{i}\right) ^{2}}{2m}}\times
\label{RateProbab1} \\
&&\func{Re}\left\{ 
\begin{array}{c}
\frac{1}{2L}\int_{-L}^{L}dx^{\prime }e^{-i\Delta q_{x}x^{\prime }}\frac{1}{2L%
}\int_{-L}^{L}dxe^{i\Delta q_{x}x}\times \\ 
\int dy^{\prime }\int dz^{\prime }\chi _{\overrightarrow{q}_{tr}^{f}}\left(
y^{\prime },z^{\prime };x^{\prime }\right) \chi _{\overrightarrow{q}%
_{tr}^{i}}^{\ast }\left( y^{\prime },z^{\prime };x^{\prime }\right) \times
\\ 
\int dy\int dz\chi _{\overrightarrow{q}_{tr}^{f}}^{\ast }\left( y,z;x\right)
\chi _{\overrightarrow{q}_{tr}^{i}}\left( y,z;x\right) \times \\ 
\int_{0}^{\infty }dte^{i\omega t}\left\langle H_{EM}^{p_{x}^{i}}\left(
x^{\prime },y^{\prime },z^{\prime };t\right) H_{EM}^{p_{x}^{f}}\left(
x,y,z;0\right) \right\rangle%
\end{array}%
\right\}  \nonumber
\end{eqnarray}%
where $\Phi $ and $A_{x}$ are the scalar and $x$-component of the
four-vector potential respectively, dominating the interaction with the
e-beam, $p_{x}=\hbar q_{x}$\ , $v=\hbar q_{x}^{i}/m$ its initial
longitudinal velocity and $m$ is the relativistic electron mass: $m=m_{0}/%
\sqrt{1-\left( v/c\right) ^{2}}$. \ In Eq.(\ref{RateProbab1}) $\Delta
q_{x}\approx \left( \omega /v\right) +\hbar \left[ \left( q_{tr}^{f}\right)
^{2}-\left( q_{tr}^{i}\right) ^{2}\right] /2mv$ is the longitudinal momentum
transfer of the e-beam, $\overrightarrow{q}_{tr}^{i}=\left(
q_{y}^{i},q_{z}^{i}\right) $ and $\overrightarrow{q}_{tr}^{f}$ its
transverse momenta, initial and final respectively, and $e^{iq_{x}x}\chi _{%
\overrightarrow{q}_{tr}}\left( y,z;x\right) $ is an unperturbed e-beam
eigenfunction with a longitudinal wave number $q_{x}$ and asymptotic
transverse wavevector $\overrightarrow{q}_{tr}$. \ Note that the width $%
\beta ^{-1}$of the Gaussian distribution function, introduced in Eq.(\ref%
{RateProbab1}) to account for the high transverse-energy cutoff caused to
the e-beam by the objective aperture, determines a region around the beam
focal plane whose length $L$ may be used as a normalization factor for the
e-beam wave functions in our model.

The condition $a^{\star }\gg c^{\star }$ ensures that the interaction
potential between the platelet and an external electron at a projected
distance $\left\vert x\right\vert $ from the platelet center decays to zero
at least as quickly as $1/x^{2}$ for $\left\vert x\right\vert >a^{\star }$
(see, e.g., Ref.\cite{Weichselbaum03}). Under these circumstances the limits
of integrations over $x$ in Eq.(\ref{RateProbab1}) may be set at $-a^{\ast }$
and $a^{\ast }$, rather than at $-L$ and $L$. \ The potential correlation
function for $t>0$ may be expressed in terms of the relevant components of
the 4-tensor photon propagator $D_{\nu ,\mu }\left( \overrightarrow{r}%
^{\prime },\overrightarrow{r};t\right) $ , $\nu ,\mu =0,1,2,3$ ($%
\leftrightarrow ct,x,y,z$), by: $-i\left\langle H_{EM}^{p_{x}^{i}}\left(
x^{\prime },y^{\prime },z^{\prime };t\right) H_{EM}^{p_{x}^{f}}\left(
x,y,z;0\right) \right\rangle =D_{0,0}\left( \overrightarrow{r}^{\prime },%
\overrightarrow{r};t\right) +\frac{p_{x}^{i}}{mc}D_{0,1}\left( 
\overrightarrow{r}^{\prime },\overrightarrow{r};t\right) +\frac{p_{x}^{f}}{mc%
}D_{1,0}\left( \overrightarrow{r}^{\prime },\overrightarrow{r};t\right) +%
\frac{p_{x}^{f}p_{x}^{i}}{\left( mc\right) ^{2}}D_{1,1}\left( 
\overrightarrow{r}^{\prime },\overrightarrow{r};t\right) $. For the sake of
simplicity we neglect the finite size effects parallel to the wide ($x-y$)
face of the platelet in the calculated photon propagator. \ Consequently the
energy loss rate in Eq.(\ref{RateProbab1}) reduces to:

\[
R\propto \int dk_{x}\int dk_{y}\func{Im}\left\{ \frac{r\left(
k_{x},k_{y},\omega \right) }{K^{\star }}\right\} 
\]

\begin{equation}
\sum_{\overrightarrow{q}_{tr}^{i},\overrightarrow{q}_{tr}^{f}}e^{-\beta
\left( \hbar q_{tr}^{i}\right) ^{2}/2m}\left\vert I\left( \overrightarrow{q}%
_{tr}^{f},\overrightarrow{q}_{tr}^{i};k_{y},K^{\star };\Delta
q_{x}-k_{x}\right) \right\vert ^{2},  \label{RateProbab}
\end{equation}%
where: 
\[
I\left( \overrightarrow{q}_{tr}^{f},\overrightarrow{q}_{tr}^{i};k_{y},K^{%
\star };\Delta q_{x}-k_{x}\right) \equiv 
\]

\begin{equation}
\frac{1}{2L}\int_{-a^{\star }}^{a^{\star }}dxe^{i\left( \Delta
q_{x}-k_{x}\right) x}J\left( \overrightarrow{q}_{tr}^{f},\overrightarrow{q}%
_{tr}^{i};k_{y},K^{\star };x\right) ,  \label{Kinematical}
\end{equation}%
and $J\left( \overrightarrow{q}_{tr}^{f},\overrightarrow{q}%
_{tr}^{i};k_{y},K^{\star };x\right) \equiv \int dze^{K^{\star }z}\int
dye^{-ik_{y}y}\chi _{\overrightarrow{q}_{tr}^{f}}^{\ast }\left( y,z;x\right)
\chi _{\overrightarrow{q}_{tr}^{i}}\left( y,z;x\right) $. \ \ The effective
surface dielectric response function, appearing in Eq.(\ref{RateProbab}), is
given by:%
\begin{eqnarray*}
r\left( k_{x},k_{y},\omega \right) &\approx &r_{0,0}+\left( \hbar
q_{x}^{i}/mc\right) r_{0,1}+ \\
&&\left( \hbar q_{x}^{f}/mc\right) r_{1,0}+\left( \hbar
^{2}q_{x}^{f}q_{x}^{i}/\left( mc\right) ^{2}\right) r_{1,1}
\end{eqnarray*}%
where $r_{\nu ,\mu }$ are the components of the EM reflection 4-tensor,
which determine the dressed photon propagator in the vacuum outside the
rectangular platelet \cite{Maniv82}:%
\[
D_{\nu ,\mu }\left( \overrightarrow{k},z,z^{\prime };\omega \right) =\frac{%
\eta _{\nu }}{2\pi K^{\ast }}\left[ \delta _{\nu ,\mu }e^{-K^{\ast
}\left\vert z-z^{\prime }\right\vert }-r_{\nu ,\mu }e^{K^{\ast }\left(
z+z^{\prime }\right) }\right] , 
\]%
\ with $\eta _{\nu }=1$ , or\ $-1$ for $\nu =0$ , or $\nu =1,2,3$
respectively.\ In the long wavelengths limit discussed in Ref.\cite{Maniv82}
we find that: 
\[
\func{Im}\left[ r\left( \overrightarrow{k},\omega \right) /K^{\star }\right]
\approx 
\]

\begin{equation}
\func{Im}\left\{ \left[ \left( K^{\star }/k^{2}\right) f_{e}+\left( \left(
v/c\right) ^{2}-\left( \omega /ck\right) ^{2}\right) f_{o}/K^{\ast }\right]
e^{-2K^{\ast }b}\right\} ,  \label{DielectResponse}
\end{equation}%
where $f_{e}=\left( \varepsilon ^{2}K^{\ast 2}-Q^{2}\right)
/D_{e}^{+}D_{e}^{-}$ , \ $f_{o}=\left( K^{\ast 2}-Q^{2}\right)
/D_{o}^{+}D_{o}^{-}$\ , $D_{e}^{+}=\varepsilon K^{\ast }+Q\tanh \left(
Qc^{\star }\right) $ , $D_{e}^{-}=\varepsilon K^{\ast }+Q\coth \left(
Qc^{\star }\right) $ , \ $D_{o}^{+}=K^{\ast }+Q\tanh \left( Qc^{\star
}\right) $ , $D_{o}^{-}=K^{\ast }+Q\coth \left( Qc^{\star }\right) $ , $Q\ =%
\sqrt{k^{2}-(\omega /c)^{2}\varepsilon \left( \omega \right) }$, and $%
\varepsilon \left( \omega \right) $ is the local bulk dielectric function of
the platelet. In the limit of a semi-infinite medium the resulting
expression reduces to the surface dielectric response function obtained in
Ref.\cite{Wang96} by using Maxwell's equations with macroscopic boundary
conditions.

The standard classical approximation for the loss function\cite{Wang96} is
obtained from Eq.(\ref{RateProbab}) by making the following assumptions: (1)
the e-beam transverse momentum distribution function $J\left( 
\overrightarrow{q}_{tr}^{f},\overrightarrow{q}_{tr}^{i};k_{y},K^{\star
};x\right) $ is a constant, that is equivalent to a $\delta $-function in
the corresponding real-space transverse coordinates, (2) the contribution of
the transverse energy to $\Delta q_{x}$ (see below Eq.(\ref{RateProbab1}))
can be neglected, and (3) the effective particle size, $a^{\star }$,
appearing as an integration limit along the beam axis, is infinite. A simple
estimate shows, however, that assumption (2) is usually invalid since the
contribution of the transverse energy to $\Delta q_{x}$ can be as large as $%
\left( \omega /v\right) $. Assumption (3), in conjunction with (1), yields
the conservation of longitudinal momentum, i.e. $\Delta q_{x}-k_{x}=0$,
which together with assumption (2) imposes the fixed condition $k_{x}=\left(
\omega /v\right) $. In the present paper we remove only the third assumption
by allowing $a^{\star }$ to be a finite length, which reflects an effective
range of the beam-particle interaction along the beam axis. \ Consequently
the longitudinal momentum distribution around $\Delta q_{x}-k_{x}=0$,
defined by the integral in Eq.(\ref{Kinematical}), is smeared and many
wavenumbers $k_{x}$ inside the light-cone start contributing to the loss
rate, Eq.(\ref{RateProbab}).

The condition for the smearing to be significant is $\pi /a^{\star }\gtrsim
\left( \omega /v\right) $, so that typically for frequencies $\omega $ in
the visible range $a^{\star }$ should be smaller than $200$ nm.
Nanoplatelets of that lengths should dramatically enhance radiative
excitations by the e-beam, previously overlooked in the literature. \ It
should be stressed that, for the sake of simplicity, this is done by
invoking the dielectric response function of an infinitely wide platelet. \
A consistent treatment of the breakdown of translation invariance is
expected, however, to further enhance all radiative channels.

As a first example we calculate the EEL function of a $100$ nm long silver
and gold platelets for an external $100$ keV e-beam at various impact
parameters (see Fig.1). To analyze the various SPP resonances one may
consider the zeros of the denominator of the extraordinary wave amplitude $%
f_{e}$ in Eq.(\ref{DielectResponse}) in the complex $k$-plane. With the
experimental optical dielectric function, $\varepsilon (\omega )$, for
silver \cite{Palik91} the resulting dispersion relation (inset, Fig.(1))
exhibits a rather flat branch of $\omega (\func{Re}k)$ inside the
light-cone, which can be attributed to radiative SPP, seen as a mirror image
of the usual non-radiative SP dispersion curve with respect to the
light-line. The sector of $\omega (\func{Re}k)$ connecting the two branches
across the light line exhibits a negative slope, where $\func{Im}k(\omega
)\propto \func{Im}\varepsilon (\omega )$ has a sharp peak. The sharp dip in
the EEL spectrum just above the classical SP frequency (at $3.8$ eV)
reflects these closely related features. These peculiar features are missing
in the loss spectrum of the gold platelet, Fig.1.

\FRAME{ftbpFU}{3.4791in}{2.8833in}{0pt}{\Qcb{EEL spectra (solid lines)\ of $%
100$ nm long Ag and Au platelets calculated for a $100$ keV external e-beam
at various impact parameters between $10$ and $40$ nm. The experimental
optical dielectric functions, $\protect\varepsilon (\protect\omega )$, for
silver and gold \protect\cite{Palik91} have been exploited. \ Dashed lines
represent spectra calculated by the classical theory. Inset: SPP dispersion
curves, $\protect\omega \left( \func{Re}k\right) ,\protect\omega \left( 
\func{Im}k\right) $ in the complex $k$-plane for silver. \ The indicated
values of $k$ and $\protect\omega $ are normalized by\ \ $k_{n}=\protect%
\omega _{n}/c,$ and $\protect\omega _{n}=10eV$, respectively. }}{}{%
flossfilmc10a50bag102040au10nmv054disprb1.eps}{\special{language "Scientific
Word";type "GRAPHIC";maintain-aspect-ratio TRUE;display "USEDEF";valid_file
"F";width 3.4791in;height 2.8833in;depth 0pt;original-width
7.8369in;original-height 6.4835in;cropleft "0";croptop "1";cropright
"1";cropbottom "0";filename
'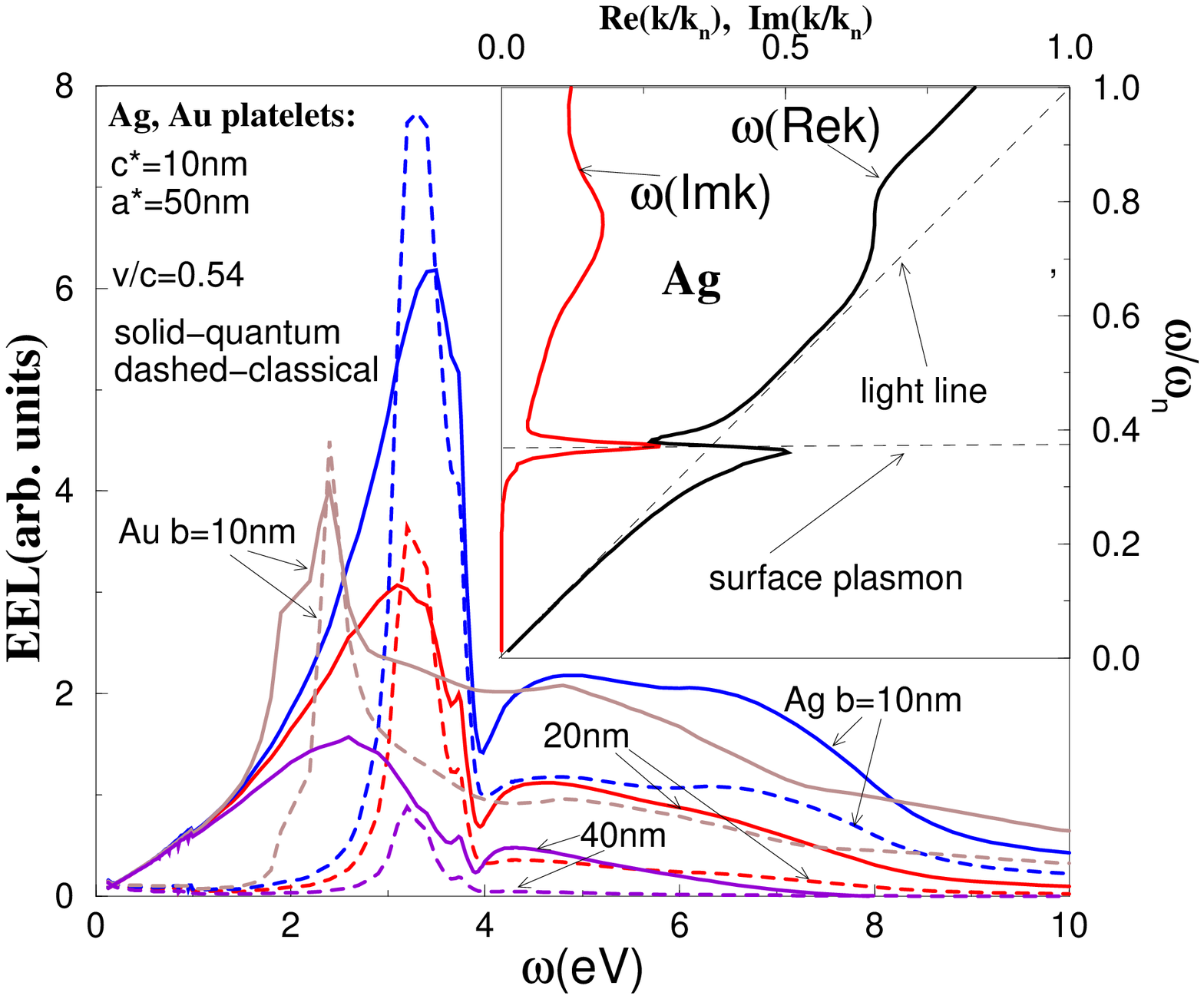';file-properties "XNPEU";}}

At slightly higher frequencies the EEL signals exhibit a pronounced rise due
to the enhanced SPP density of states associated with the flat radiative SPP
branch. The corresponding EEL peak intensity exhibits a remarkably weak
attenuation, in marked contrast to the exponential attenuation of the main
SP peak, calculated in the classical limit for increasing impact parameter.
The radiative nature of the beam-particle coupling shown in Fig.(1) is even
more pronounced in the low energy region below the main SP peak, where the
classically calculated signal drops to zero. Here our calculated EEL
function exhibits a pronounced broad band with linearly increasing intensity
for increasing frequency and almost no attenuation with increasing impact
parameter. These features are due to the fact that the loss signal well
below the main SP frequency is dominated by the contribution from the
ordinary wave amplitude $f_{o}$, appearing in Eq.(\ref{DielectResponse}),
which is singularly enhanced near the light line (where $K^{\ast
}\rightarrow 0$ ), and thus reflecting the nearly pure (transverse) photonic
nature of the excitations by the e-beam in this 'classically forbidden'
region.

\FRAME{ftbpFU}{3.5423in}{2.9352in}{0pt}{\Qcb{EEL spectra (solid lines) of a $%
100$ keV e-beam propagating parallel to the $x$-axis of a rectangular SiO$%
_{2}$ platelet at distances $b=10,20,30$ nm above its wide ($x-y$) face. \
The platelet half sides along the $x$, and $z$ axes are: $a^{\star }=50$ nm,
and $c^{\star }=5$ nm respectively. The corresponding spectra (dashed lines)
obtained from the classical theory are also shown for comparison. \ Note the
close similarity of the classical spectrum for $b=10$ nm to that obtained in
Ref.\protect\cite{deAbajo02} for the same e-beam velocity at nearly the same
impact parameter parallel to a sharp wedge (see inset). Inset: A schematic
illustration of the SiO$_{2}$ platelet showing the e-beam (wide face)
configuration studied in the present paper and the (wedge) configuration
investigated in Ref.\protect\cite{deAbajo02}. }}{}{%
fsio2fc5a50b102030v054lossinsetprlep.eps}{\special{language "Scientific
Word";type "GRAPHIC";maintain-aspect-ratio TRUE;display "USEDEF";valid_file
"F";width 3.5423in;height 2.9352in;depth 0pt;original-width
7.8369in;original-height 6.4835in;cropleft "0";croptop "1";cropright
"1";cropbottom "0";filename
'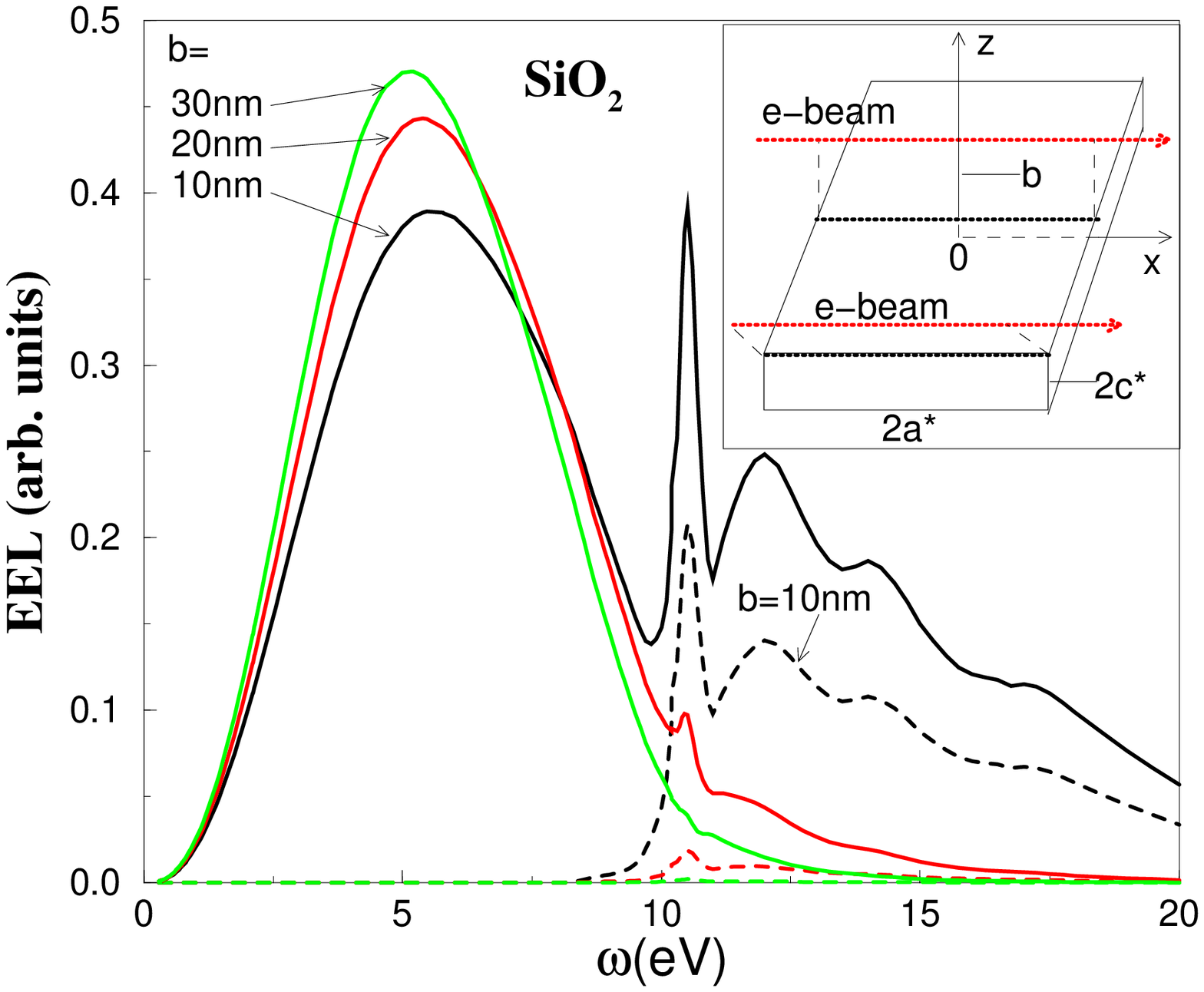';file-properties "XNPEU";}}

A similar and even more dramatic situation arises in the forbidden energy
gap of the electron-hole pair excitations in semiconductors and insulators,
as shown, e.g. in Fig.(2) for a $100$ nm long SiO$_{2}$ platelet. The EEL
spectra of an external $100$ keV e-beam, propagating parallel to the $x-y$
face of the platelet at various impact parameters $b$, reveal a pronounced
broad peak within the forbidden gap region, which does not decay with
increasing $b$ values. Similarly to the situation with the silver platelet
well below the main SP peak, the strong radiative nature of this feature
arises from the ordinary wave amplitude $f_{o\text{ }}$, corresponding to
the excitation of purely transverse EM waves, polarized within the $x-y$
plane, which totally dominates the loss signal in the gap region.

The spectrum shown in Fig.(2) for an impact parameter $b=10$ nm may be
compared to the result reported in Ref.\cite{deAbajo02} for an electron
moving parallel to a $90^{\circ }$ SiO$_{2}$ wedge at a distance of $8.5$ nm
(see Inset to Fig.(2)). The pronounced radiative broad band within the gap
region, obtained in our calculation, dramatically contrasts the vanishing
loss signal shown there in Fig.(4) for an electron with the same velocity ( $%
v=0.54c$) and nearly the same impact parameter. The lack of far-field
coupling in the latter theoretical approach restricted the fast external
e-beam to excite EM waves propagating within the dielectric medium only\cite%
{Zabala03}, similar to ordinary waveguide modes which can develop within a
thin SiO$_{2}$ slab in the forbidden gap region where $\func{Re}\varepsilon
\left( \omega \right) \approx 2$ , and $\func{Im}\varepsilon \left( \omega
\right) \rightarrow 0$. \ For ideal planar geometry (as assumed in our
calculation of the dielectric response function $r\left( \overrightarrow{k}%
,\omega \right) $), the corresponding waveguide modes appear as extremely
narrow resonances which can not be excited by an e-beam with $\Delta q_{x}$
values outside the light cone due to the vanishingly small dielectric
damping, $\func{Im}\varepsilon \left( \omega \right) $.

Such radiation excitations become possible for the nonplanar geometries
studied in Refs.\cite{Zabala03},\cite{deAbajo02} even under the recoilless
scattering approximation exploited there (but only above a threshold beam
energy considerably higher than $100$ keV) due to the translational
symmetry-broken dielectric media considered in their calculations. Yet, the
corresponding Cherenkov channels remain fundamentally different from the
ones we propose: \ The opening of scattering channels with wave numbers
inside the light cone allows coupling of the e-beam to the continuum of EM
modes which are extended into the vacuum, leading to dramatic enhancement of
the loss signal at e-beam energies around $100$ keV. The relative strength
of the radiative effect calculated here may be further appreciated by noting
the close proximity of the e-beam to the sharp SiO$_{2}$ wedge in \cite%
{deAbajo02},which strongly enhances the (evanescent) near-field coupling
with the e-beam.

In summary, applying a quantum-mechanical approach to the scattering
problem, we have shown that Cherenkov radiation of highly focused
relativistic e-beams in STEM, discussed recently in the literature\cite%
{Zabala03,deAbajo98-99,Ochiai04}, have a much broader scope than originally
presented. The dramatic enhancement of radiative channels, imposed by the
finite nanoparticle size, arises from the breakdown of momentum conservation
in the inelastic scattering event along the e-beam axis. Further enhancement
should be realized due to the extreme lateral confinement of the e-beam and
its associated transverse momentum uncertainty, which were neglected here.
The radiation predicted to be emitted from both conducting and insulating
nanoplatelets can be generated at impact parameters much larger than the
evanescent tail of the excited surface EM modes. Large deviations from the
classical, nonradiative EEL signal are found to persist also at small impact
parameters, which can be readily tested experimentally. The proposed
spectroscopic experiments are closely related to recent developments in the
field of SPP-based far-field optics.\cite{Lezec02}. Far field EELS can
therefore become a useful complementary tool for an already recognized new
emerging field.

We would like to thank Boris Lembrikov for helpful discussions. This
research was supported by SENIEL\ OSTROW\ RESEARCH\ FUND, and by the fund
for the promotion of research at the Technion.

. \qquad \qquad \qquad \qquad \qquad \qquad \qquad \qquad \qquad \qquad
\qquad \qquad \qquad \qquad \qquad \qquad \qquad \qquad \qquad \qquad \qquad
\qquad \qquad \qquad \qquad \qquad \qquad \qquad \qquad \qquad \qquad \qquad
\qquad \qquad \qquad \qquad \qquad \qquad \qquad \qquad \qquad \qquad \qquad
\qquad \qquad \qquad \qquad \qquad \qquad \qquad \qquad \qquad \qquad \qquad
\qquad \qquad \qquad \qquad \qquad \qquad \qquad \qquad \qquad \qquad \qquad
\qquad\


\begin{thebibliography}{99}
\bibitem{Batson83} P. E. Batson, Ultramicroscopy \textbf{11}, 299 (1983).

\bibitem{Howie85} A. Howie, R.H.Milne, Ultramicroscopy 18, 427-434 (1985).
P.M. Echenique, A. Howie, \textit{ibid.} 16, 269-272 (1985).

\bibitem{Muller93} D. A. Muller, W. Tsou, R. Raj, and J. Silcox, Nature
(London) \textbf{366}, 725 (1993).

\bibitem{Cohen98} H. Cohen, T. Maniv, R. Tenne, Y. Rosenfeld Hacohen, O.
Stephan, and C. Colliex, Phys. Rev. Lett. \textbf{80}, 782 (1998); P.M.
Echenique, A. Howie and R.H. Ritchie, \textit{ibid.} \textbf{83}, 658
(1999); H. Cohen \textit{et al}.,\textit{ibid.} \textbf{83}, 659 (1999).

\bibitem{Lembrikov03} B. I. Lembrikov, M. A. Itskovsky, H. Cohen, and \ T.
Maniv, Phys. Rev. B \textbf{67, }085401, 2003.

\bibitem{Cohen03} H. Cohen, B. I. Lembrikov, M. A. Itskovsky, and T. Maniv,
NANOLETTERS \textbf{3} (2), 203 (2003).

\bibitem{Raether80} H. Raether, \textit{Excitation of Plasmons and Interband
Transitions by electrons} (Springer, New York, 1980).

\bibitem{Zabala03} N. Zabala, A. G. Pattantyus-Abraham, A. Rivacoba, F. J.
Garcia de Abajo, and M. O. Wolf, Phys. Rev. B \textbf{68}, 245407 (2003).

\bibitem{deAbajo98-99} F. J. Garcia de Abajo and A. Howie, Phys. Rev. Lett. 
\textbf{80}, 5180 (1998); F. J. Garcia de Abajo, Phys. Rev. B \textbf{59} ,
3095 (1999).

\bibitem{Ochiai04} Tetsuyuki Ochiai and Kazuo Ohtaka, Phys. Rev. B \textbf{69%
}, 125106 (2004).

\bibitem{deAbajo03} F. J. Garcia de Abajo, A. G. Pattantyus-Abraham, N.
Zabala, A. Rivacoba, M. O. Wolf, P. M. Echenique, Phys. Rev. Lett. \textbf{91%
}, 143902 (2003).

\bibitem{Maniv82} T. Maniv and H. Metiu, J. Chem. Phys. \textbf{76} , 696
(1982); T. Maniv, Phys. Rev. B \textbf{26, }2856 (1982).

\bibitem{Weichselbaum03} A. Weichselbaum and S. E. Ulloa, Phys. Rev. B 
\textbf{68}, 056707 (2003).

\bibitem{Wang96} Z. I. Wang, MICRON \textbf{27}, 265-299 (1996).

\bibitem{Palik91} \textit{Handbook of Optical Constants of Solids II} ,
edited by Edward\ D. Palik (Academic Press, Boston, 1991).

\bibitem{deAbajo02} F. J. Garcia de Abajo and A. Howie, Phys. Rev. B \textbf{%
65}, 115418 (2002).

\bibitem{Lezec02} H. J. Lezec, A. Degiron, E. Devaux, R. A. Linke, L.
Martin-Moreno, F. J. Garcia-Vidal, T. W. Ebbesen, Science \textbf{297}, 820
(2002).
\end{thebibliography}
\end{document}